\documentclass[journal]{IEEEtran}
\usepackage{graphicx}
\usepackage{amsmath}
\usepackage{amsthm}
\usepackage{amssymb}
\usepackage{import}
\usepackage{color}
\usepackage{array}
\usepackage{diagbox}
\usepackage{multicol}
\usepackage{multirow}
\usepackage{placeins}
\usepackage[linesnumbered,ruled]{algorithm2e}
\usepackage{hyperref}
\usepackage{balance}
\usepackage[normalem]{ ulem }
\usepackage{soul}
\usepackage{mathtools, nccmath}
\usepackage{caption}
\usepackage{subcaption}
\usepackage{tabularx}
\usepackage{placeins}
\usepackage{cite}
\title{A mobile observer method for the estimation of road traffic using communicating vehicles}

\author{
\IEEEauthorblockN{Cyril Nguyen Van Phu}\\ 
\IEEEauthorblockA{COSYS-GRETTIA, Univ Gustave Eiffel, IFSTTAR, F-77454 Marne-la-Vall\'ee, France} \\
}

\newtheorem{prop}{Proposition}

\begin{document}
\maketitle
\begin{abstract}
Estimation of road traffic is a fundamental problem which has been addressed with a variety of methods.
In the present paper, a variant of the mobile observer method is proposed.
It is assumed that some vehicles composing the road traffic are communicating vehicles.
These communicating vehicles broadcast periodically beacon messages.
The proposed method uses only these beacon messages as input data, and needs no additional equipment such as radar
or GPS device in order to estimate the road traffic. An estimator of the penetration ratio of communicating vehicles is derived.
The model is tested with the bi-directional simulation framework VEINS, which combines
a microscopic road traffic simulator and a communication simulator.
The preliminary results show the potential of the method and confirm the validity of the approach.
\end{abstract}


\section{Introduction and state of the art}
\subsection{Introduction}
\label{section:intro}
Mobility of goods and individuals is central to modern societies, where
vehicular road traffic is one of the main transportation mode.
When the road traffic demand for transportation exceeds the road traffic capacity, congestion occurs.
Road traffic congestion is a highlighted problem because it leads to major consequences on the environment, the economy, the health of the individuals, and the lost time~\cite{simeonova2021congestion}.
Road traffic control strategies have been proposed in order to reduce congestion and to improve road traffic conditions.
Prior to applying control strategies, there is the need to have an accurate estimation of the road traffic.
Actually, the emergence of communicating vehicles offer new possibilities for the estimation of road traffic.
In~\cite{GUO2019313}, the authors give a survey of control and estimation of road traffic methods with communicating or automated vehicles.

Concerning the estimation of road traffic, various approaches have been proposed.
In 1954, Wardrop and Charlesworth~\cite{doi:10.1680/ipeds.1954.11628} have proposed a moving observer method for the estimation of road traffic.
In 2017, Florin and Olariu~\cite{7534864} have proposed a variant of the mobile observer method, using communicating vehicles, and radar equipments.
In the present paper, it is shown that the mobile observer method can be used with only communication messages, releasing the need for radar or even GPS localization.
The outline of the present papers is as follows :
In section~\ref{subsection:state-art}, the literature context for the actual mobile observer method is given.
In section~\ref{original_method}, the historical method is formulated.
In section~\ref{model}, the proposed variant model along with its contributions are given.
In section~\ref{simulation}, simulations are performed in order to assess that the method is functional, and to provide the preliminary results.
Finally, the conclusion~\ref{conclusion} opens perspectives to the present work.

\subsection{State of the art}
\label{subsection:state-art}
Using networking technologies, the Mobile Ad Hoc Networks (MANETs) concept has been proposed.
A MANET is a wireless communication network composed of mobile nodes, which do not rely on a central infrastructure in order to function.
Vehicular Ad Hoc Networks (VANETs) are a type of Mobile Ad Hoc Networks (MANETs) where the mobile nodes are moving vehicles.

Lee and Atkison~\cite{LEE2021100310} give a survey of VANETs principles and applications.
VANETs are characterized by the high number of communicating nodes in the network as well as the high speed of the nodes (the vehicles) compared to the general case 
of Mobile Ad Hoc Networks (MANETs). Indeed, VANETs ``are one of the fastest-moving environments where modern technologies are currently employing computer devices''.
VANETs can transmit data at high speed, that being required for safety applications.
VANETs can provide information on the location of the vehicles, which induce potential beneficial applications as well as privacy and security concerns.
The high mobility of nodes in VANETs makes the multi hop routing of messages over long distances challenging.

Anwer and Guy~\cite{anwer2014survey} give a survey of wireless communication protocols and technologies used in VANETs.
The authors describe the key technologies standards such as ``802.11p, P1609 protocols, Cellular System, CALM, MBWA, WiMAX, Microwave, Bluetooth and ZigBee'' and how they can be used in VANETs according to their characteristics such as 
the data transmission rate, the signal coverage, the security and other parameters.

Darwish and Abu Bakar~\cite{darwish_traffic_2015} give a bibliographic survey on traffic density estimation in vehicular ad hoc networks.
In their paper, the state of the art of the estimation of road traffic density with VANETs is presented, either with or without communication infrastructures, such as
Road Side Units (RSU). In their paper, the authors group infrastructure free vehicle density estimation methods in three sets which respectively use
statistical methods, VANETs communication and traffic flow information, or clustering and counting.

Within VANETs, Mao and Mao\cite{6555328} use the number of neighbours in the vehicular communication network in order to estimate the road traffic density.
They consider that the vehicles are distributed along the road segment according to a Poisson probability distribution.
By using the number of neighbours in each vehicular network hop, a maximum likelihood estimator of the road traffic density is given.
Derrmann et al.~\cite{8317718} use mobile data available to the mobile networks operators and especially signaling data, 
which is a ``meta-data from the mobile network infrastructure that is generated e.g. if a user initiates a call or moves from one antenna to another (a so-called handover). ''
in order to estimate urban road traffic states. They compare their estimation of road traffic states with a reference data set provided by floating car data (FCD).
Valerio et al.~\cite{5073548} survey the various approaches for estimating road traffic with cellular communication networks.
They distinguish between active techniques which induce additional communication traffic and passive techniques which collect data without impacting the communication network load.
In addition, the authors propose a framework for the gathering of cellular communication data with the objective of road traffic estimation.

In 2007, Artimy~\cite{artimy_local_2007} gives a vehicles density estimator which uses the two fluid model and a car following model.
It is applied to design a transmission range algorithm for VANETs. The input data of its estimator is the stop time of vehicles related to the total travel time on the considered road section. The road density estimator is evaluated in simulation as well as the maximum transmission range assignment method.

Another approach by Yu et al.~\cite{yu_vanet_2013} uses VANETs communication and beacon messages in order for a vehicle to compute the road traffic density of vehicles on the road 
to which it belongs. In VANETs, the topology of the communication network is changing rapidly due to the high speed of the mobile communication nodes. This vehicle density estimation algorithm is useful for conceiving packet routing communication protocols. Such protocols are essential to route communication packets in the mobile ad-hoc network. Of course, 
the density of vehicles on a road is also an information useful for road traffic management and control. In their paper, the authors do not evaluate the performances
of their road vehicle density estimation algorithm.

Concerning road traffic estimation on highways, Seo et al.~\cite{seo_traffic_2017} give a survey on traffic state estimation on highway. The authors present traffic flow models and data used for traffic state estimation,
which form the basis for various traffic state estimation approaches.
Among these approaches, Wardrop and Charlesworth~\cite{doi:10.1680/ipeds.1954.11628} have proposed a method for estimating the road traffic speed and flow by using a mobile observer.
The mobile observer is a car moving in the traffic and which observes ``the number of vehicles met in the section when travelling against
the stream, the number of vehicles that overtake the observer minus the number of vehicles
he overtakes, the journey time of the observer when travelling against the stream, and the journey time of the observer when travelling with the stream.''.
The method consists in a car measuring these input data and performing various runs on a road section, in the direction with and against the stream.
From these input data provided by the moving car, the authors derive expressions for the traffic flow and space mean speed of the considered road section.

In 2012, Bauza and Gozalves~\cite{bauza_traffic_2013} have proposed an estimator for the density of vehicles on road, 
combined with fuzzy logic and communication protocols, for the detection of the congestions occuring in a road network.

In 2017, Florin and Olariu~\cite{7534864} have proposed a variant of the mobile observer method by Wardrop and Charlesworth~\cite{doi:10.1680/ipeds.1954.11628}.
They assume that the vehicles are equipped with short and long range radars, as well as communication equipments.
Their variant method uses only the tally which is ``the number of vehicles they pass and the number of
vehicles that pass them in a given road segment.''
The system of equations proposed by Wardrop and Charlesworth needs two equations in order to be solved.
Instead of using the count of vehicles encountered against the stream as a second equation like in the original method, 
the authors use inter-vehicular communication and communicate the additional equation from another equipped vehicle moving at a different speed nearby in the traffic.
The authors compute the density, average speed and flow of road traffic in real time.
They have evaluated their method by performing simulations with ns3 simulator~\cite{henderson2008network} for various equipment penetration ratios.

In a subsequent paper~\cite{9807677}, Florin and Olariu use the tallies for vehicles moving in opposite directions.
They also perform simulations with SUMO microscopic road traffic simulator~\cite{krajzewicz_recent_2012}.

In 2020, Florin and Olariu~\cite{florin_towards_2020} have proposed a method for the estimation of traffic state using a moving observer method. They still use tallies as input data for the estimation of traffic density, and the information is disseminated using V2V wireless communications.

In 2021, Kuwahara et al.~\cite{kuwahara_traffic_2021} have published a method for the estimation of road traffic, using backward probe vehicles, i.e. probe vehicles moving in the opposite direction to
the traffic which is estimated. The backward probe vehicles measure the passing times of vehicles running forward when it meets them along its trajectory.
As an application of their method, the authors also estimate the road traffic in a scenario of incident.
It is noticeable that the authors do not explicit the method for measuring time headways of vehicles running forward, that the backward probe vehicles encounter.

Van Erp et al.~\cite{VANERP2018281} combine stationary and moving observers data inputs to estimate macroscopic traffic flow parameters in homogeneous and stationary conditions.
Given the cumulative number of vehicles at three points in the space time domain provided by moving and static observers,
the authors show that the traffic conditions can be estimated for the full space-time domain.

\section{Problem statement}
\label{original_method}
\subsection{Problem statement}
\label{subsection:problem}
The current section gives the notations used in the paper,
and the formulation of the mobile observer method proposed by Wardrop and Charlesworth~\cite{doi:10.1680/ipeds.1954.11628} in 1954.
\begin{table}[htbp]
\begin{tabular}{|l|p{6cm}|}
\hline
Name & Definition \\
\hline
$l$ & the length of the road section. \\
$y_i$ & the number of beacon messages received by a mobile observer from a vehicle $i$. \\
$q_i$ & the flow of vehicles moving at speed $v_i$ in the direction $a$.\\
$v_i$ & the speed of the vehicles composing the flow $q_i$ in the direction $a$. \\
$q=\sum_i q_i$ & the total flow of vehicles moving in the direction $a$.\\
$q_{1i}$ & the flow of vehicles moving at speed $v_i$ in the direction $a$ relative to an observer $1$ moving in the direction $a$.\\
$q_{2i}$ & the flow of vehicles moving at speed $v_i$ in the direction $a$ relative to an observer $2$ moving in the direction $w$.\\
$v_{1}$ & the speed of the moving observer $1$ moving in the direction $a$.\\
$v_{2}$ & the speed of the moving observer $2$ moving in the direction $w$.\\
$t_{1}$ & the travel time to cross road section of length $l$ for the observer $1$ moving in the direction $a$.\\
$t_{2}$ & the travel time to cross road section of length $l$ for the observer $2$ moving in the direction $w$.\\
$x_{1}$ & the number of vehicles moving in direction $a$, passed by, or which pass the moving observer $1$ moving in the direction $a$.\\
$x_{2}$ & the number of vehicles moving in direction $a$ encountered by the moving observer $2$ moving in the direction $w$.\\
$x_{1i}$ & the number of vehicles moving at speed $v_i$ in the direction $a$ passed by, or which pass observer $1$ moving in the direction $a$ on the road section of length $l$.\\
$x_{2i}$ & the number of vehicles moving at speed $v_i$ in the direction $a$ which encounter observer $2$ moving in the direction $w$ on the road section of length $l$.\\
$N_{1}$ & a random variable which is the number of communicating vehicles moving in the direction $a$ passed by, or which pass observer $1$ moving in the direction $a$ on the road section of length $l$.\\
$N_{2}$ & a random variable which is the number of communicating vehicles moving in the direction $a$ which encounter observer $2$ moving in the direction $w$ on the road section of length $l$.\\
$m_1$ & the number of communicating vehicles moving in the same direction that the moving observer 1, and in its radio range $s$, at a time $t$.\\
$n_1$ & the number of communicating vehicles moving in the same direction that the moving observer 1, which have entered the radio range of moving observer 1 during the travel time.\\
$n_2$ & the number of communicating vehicles moving in the opposite direction that the moving observer 1, which have entered the radio range of moving observer 1 during the travel time.\\
$p$ & the penetration ratio of the communicating vehicles. The penetration ratio is determined by other methods : see for example Nguyen Van Phu and Farhi~\cite{9864074} or Comert~\cite{comert_queue_2016}.\\
\hline
\end{tabular}
\caption{Notations}
\label{tab:notations}
\end{table}

The flow of vehicles $q_i$ flowing in the direction $a$ relative to the moving observer $1$ moving in the direction $a$ is given by :
\begin{equation}
q_{1i}=q_i(1-v_1/v_i) \nonumber
\end{equation}

The number of vehicles passed by or which pass observer $1$, and which are moving in the same direction $a$ at speed $v_i$ is given by :
\begin{equation}
x_{1i}=q_{1i}t_1 \nonumber
\end{equation}

\begin{equation}
\label{eq:x_1i}
x_{1i}=q_i(t_1-l/v_i)\\
\end{equation}


The flow of vehicles $q_i$ flowing in the direction $a$ relative to the moving observer $2$ moving in the direction $w$ is given by :
\begin{equation}
q_{2i}=q_i(1+v_2/v_i) \nonumber
\end{equation}

The number of vehicles encountered by the observer $2$, and which are moving in the direction $a$ at speed $v_i$ is given by :
\begin{equation}
x_{2i}=q_{2i}t_2 \nonumber
\end{equation}

\begin{equation}
\label{eq:x_2i}
x_{2i}=q_i(t_2+l/v_i) \nonumber
\end{equation}

Summing Equation~(\ref{eq:x_1i}) over all flows $q_i$ moving at speeds $v_i$ in direction $a$ gives : 
\begin{equation}
x_1=\sum_i x_{1i}=\sum_i q_i(t_1-l/v_i)=\sum_i q_i(t_1-t_i) \nonumber
\end{equation}
\begin{equation}
x_1=qt_1-\sum_i q_it_i=q(t_1-\frac{\sum_i q_it_i}{q})\nonumber
\end{equation}
Let us introduce $\bar{t}$ the average travel time according to a distribution with frequencies $q_i/q$ : 

\begin{equation}
\bar{t}=\sum_i \frac{q_i}{q}t_i \nonumber
\end{equation}

We have :
\begin{equation}
\label{eq:x_1}
x_1=q(t_1-\bar{t})
\end{equation}
Similarly, 
\begin{equation}
\label{eq:x_2}
x_2=q(t_2+\bar{t})
\end{equation}

The system of Equations~(\ref{eq:x_1})~(\ref{eq:x_2}) can be solved to determine the two unknowns $q$ and $\bar{t}$.
The fundamental traffic parameters are related with $$q=\kappa v_{avg}$$ where $q$ is the flow, $\kappa$ is the density and $v_{avg}=l/\bar{t}$ is the space mean speed of vehicles~\cite{wardrop_road_1952}.
Knowing the two variables $q$ and $v_{avg}$ fully determine the third one, the density of vehicles $\kappa$.


\section{Model}
\label{model}

\subsection{Assumptions}
The present paper is using the following assumptions :
\begin{itemize}
 \item some vehicles composing the road traffic are communicating vehicles which send periodically beacon messages, according 
to the WAVE communication protocol~\cite{6755433}.
\item the penetration ratio $p$ of these communicating vehicles is variable.
\item  the distribution of the speeds of the vehicles is such that for all vehicles $j$ of the same stream that the mobile observer $\exists M, |v_1-v_j| \leq M$, where $v_1$ is the speed of the moving observer. It is assumed that for all vehicles $i$ of the opposite stream $v_1+v_i>M$, such that the messages of the two directions can be differentiated. This means that the speeds of the vehicles should not be too low.
\end{itemize}

\subsection{Contributions}
The mobile observer method proposed by Wardrop and Charlesworth~\cite{doi:10.1680/ipeds.1954.11628} was relying on counting and board papers.
Recent works have investigated how to use the mobile observer method with modern measurement techniques.
These methods rely either on radar equipment~\cite{7534864} or do not specify the method for measurement~\cite{kuwahara_traffic_2021}.
While some methods for the estimation of vehicles density on a road using VANETs have been proposed as presented in the bibliographic survey by Darwish and Abu Bakar~\cite{darwish_traffic_2015},
most of these methods were designed in order to provide functional routing of messages in VANETs and very few of these algorithms have been evaluated with experiments, either numerical
or in the field.
The main contribution of the present paper consists in explaining how the historical mobile observer method
can be applied using communication messages only in order to estimate the road traffic parameters in real time.
The applications may be routing VANETs messages, selecting the transmission range of vehicles in relation the density of vehicles on the road network,
or road traffic management and control.
The present paper performs realistic numerical experiments which simulate the WAVE communication protocol~\cite{6755433} 
using VEINS framework~\cite{sommer2011bidirectionally}
which bi-directionally couples a microscopic road traffic simulator SUMO~\cite{krajzewicz_recent_2012} and a communication simulator OMNET++~\cite{varga_overview_2010}.
One additional contribution of this paper is that the method is not limited when the flow of vehicles is low or moderate, which is a common problem in the estimation of road traffic using floating car data and fundamental diagrams~\cite{turner2017using}.
Finally, the derivation of a penetration ratio estimator for the communicating vehicles is given and its performances are evaluated in simulation.

\subsection{Differentiating messages}
\label{subsection:messages}
The present work assumes the possibility of counting the vehicles encountered, by using Basic Safety Messages (BSM) from communicating vehicles.
BSM messages are part of the WAVE communication protocol which enables inter-vehicular communication.
BSM messages are sent typically at a frequency $f=10$ Hz~\cite{LIU201683}.
The moving observer receives BSM from communicating vehicles moving against and with the stream.
The two types of communicating vehicles (against and with the stream) can be differentiated by the properties of the BSM messages sent.
If a communicating vehicle is moving against the moving observer, then the connection will be very brief with few BSM messages received.
On the other hand, if a communicating vehicle is moving in the same direction than the moving observer, then the connection will last longer and more BSM messages will be received.
This is because the radio range which enables the reception of BSM messages decreases faster as vehicles move in opposite directions than if
vehicles are moving the same direction (in this case, the vehicles stay close longer). 
In this paper, the differentiation of communicating vehicles moving with or against the direction of the moving observer is investigated.

Let us consider the time windows $t_a,t_w$ during which the moving observer and a vehicle are communicating in the radio range $s$, respectively for vehicles 
$i$ moving in the opposite direction or vehicles $j$ moving in the same direction that the mobile observer.
Concerning vehicles $i$ moving in the opposite direction of the mobile observer, the time window is : 
\begin{equation}
 t_a=\frac{2s}{v_1+v_i}
 \label{eq:ta}
\end{equation}
We notice that there is a factor $2$ at the numerator, because the time window is composed of the time period before the two vehicles have intersected, plus the time period after the two vehicles have intersected. Each time period equals $s/(v_1+v_i)$.
For the vehicles $j$ moving in the same direction that the moving observer, we have for $v_1\neq v_j$ :
\begin{equation}
 t_w=\frac{2s}{|v_1-v_j|}
 \label{eq:tw}
\end{equation}
For vehicles $j$ of the same stream that the mobile observer, $\exists M=\max\limits_{j}\{v_1-v_j\}$ such that $\forall j, |v_1-v_j|\leq M$.
For vehicles $i$ of the opposite stream of the mobile observer, if $v_1+v_i > M$, then $v_1+v_i>|v_1-v_j|$, and it is expected that $t_a<t_w$.
The number of BSM messages received during $t_a$ and $t_w$ are different, which allows to discern between the messages from the 
streams of the two directions.

Each communicating vehicle keeps a record with the number of beacon messages received from each other communicating vehicle.
Such a record is a vector where each coefficient is the number of beacons sent from a given communicating vehicle.
Then, for each vehicle, the vector is divided in two sets : the set of beacons received from communicating vehicles moving in the same direction, and
the set of beacons received from communicating vehicles moving in the opposite direction.

The proposed procedure for clustering the messages in two sets consists in :
\begin{itemize}
 \item given $v_1$, $v_i$ and $v_j$ communicated in the messages, and an estimation of the radio range $s$, compute $t_a$ and $t_w$ according to Equations~(\ref{eq:ta})~and~(\ref{eq:tw}).
 \item given $y_i$ the number of messages received by the mobile observer from vehicle $i$ and $f$ the frequency of messages emission :
 \item if $|y_i-ft_a|<|y_i-ft_w|$, then vehicle $i$ is moving against the mobile observer.
 \item Otherwise, vehicle $i$ is moving with the stream 
 of the mobile observer.
\end{itemize}
The result of the clustering is the number of communicating vehicles encountered by the mobile observer, moving either
in the same direction $n_1$, or in the opposite direction $n_2$.

\subsection{Estimation of road traffic}
\label{subsection:model}
It is assumed that the flow of vehicles is constant over the radio range of the moving observer, and thus does not vary if the point of measurement is
the boundary of the radio range of the moving observer, or the moving observer itself.
With this hypothesis, the quantity of vehicles $x_2$ which pass, or are passed by the moving observer is equivalent
to the number of vehicles which pass or are passed by the boundary of the radio range of the moving observer.
Indeed, $x_2$ is the number of vehicles entering the radio range of the moving observer, whether these vehicles are communicating vehicles, or not.
If the flow of vehicles is not constant on the considered radio range, then $x_1$ and $x_2$ should be computed with a more sophisticated model.
However, the assumption that the flow is in a steady state is very realistic.

$n_2$ is the number of communicating vehicles entering the radio range of the moving observer.
The probability that $n_2$ communicating vehicles are encountered in the on coming traffic is following a binomial probability distribution,
where $n_2$ is measured and $p$ is assumed to be known :
\begin{equation}
P(N_2=n_2)=\binom{x_2}{n_2}p^{n_2}(1-p)^{x_2-n_2}
\end{equation}
Then, $x_2$ is determined with maximum likelihood estimation~\cite{myung_tutorial_2003} :
\begin{equation}
\label{eq:likelihood1}
x_2=\arg\max \binom{x_2}{n_2}p^{n_2}(1-p)^{x_2-n_2}
\end{equation}



Concerning Equation~(\ref{eq:likelihood1}), the binomial probability distribution is unimodal~\cite{binomial_rundel}.
Let us define $r:=P(N_2=n_2+1)/P(N_2=n_2)$.
The maximum of the binomial probability distribution is reached when $r$ switches from being greater than $1$ to being less than $1$.
$r\leq1$ is equivalent to :
\begin{equation}
\label{eq:binomial_max}
n_2\geq x_2p-1+p
\end{equation}
The maximum probability is reached for $x_2p-1+p$ being the maximum value less than $n_2$ :
\begin{equation}
\label{eq:binomial_max2}
x_2=\lfloor \frac{n_2+1-p}{p} \rfloor 
\end{equation}

Similarly to $x_2$, $x_1$ is given by :

\begin{equation}
\label{eq:binomial_max1}
x_1=\lfloor \frac{n_1+1-p}{p} \rfloor 
\end{equation}

Then, by combining Equation~(\ref{eq:x_1})~and~(\ref{eq:x_2}), we deduce the flow :
\begin{equation}
\label{eq:q}
q=\frac{x_1+x_2}{t_1+t_2}
\end{equation}

$\bar{t}$ is measured directly by the communicating vehicles.
The density is given by :
\begin{equation}
\label{eq:density3}
\kappa=q\bar{t}/l
\end{equation}

\begin{prop}
\label{prop:density1}
\begin{equation}
\kappa=\frac{(x_1+x_2)\bar{t}}{(t_1+t_2)l}
\end{equation}
\end{prop}

Another approach to calculate the density of vehicles on the road consists in counting the number of vehicles and dividing it by the length of the considered area~\cite{bauza_traffic_2013}.
This kind of approach is using GPS localization.
%

Rather than using GPS localization to have a reference length of road, it is proposed to use the radio range $2s$, which simplifies the equipments needed.
$2s$ is given by Equations~(\ref{eq:ta})~(\ref{eq:tw}) :
%
\begin{equation}
\label{eq:s1}
2s=|v_1-v_j|y_j/f
\end{equation}

\begin{equation}
\label{eq:s2}
2s=(v_1+v_i)y_i/f
\end{equation}

Let us denote $m_1$ the number of communicating vehicles moving in the same direction that a moving observer, and in its radio range $s$, at a time $t$.
The difference between $m_1$ and $n_1$ is that $n_1$ is the total number of communicating vehicles moving in the same direction and which have entered the radio range of the moving observer
over the travel time.

\begin{prop}
\label{prop:density2}
\begin{equation}
\kappa=\frac{fn_2(m_1+1-p)}{p\sum_{i=1..n_2}{y_i(v_1+v_i)}}
\end{equation}
\end{prop}
\proof
By denoting $\bar{s}$ the average value of the radio range $s$, we have :
\begin{equation}
 \kappa=\frac{m_1+1-p}{2p\bar{s}}
\end{equation}
$2\bar{s}$ is derived from Equation~(\ref{eq:s1})~(\ref{eq:s2}):
\begin{equation}
\label{eq:s}
2\bar{s}=\sum_{i=1..n2}y_i(v_1+v_i)/fn_2
\end{equation}
\endproof


\subsection{Penetration ratio estimator}
Let us assume that the flow of vehicles $q$ is known by another method, like a static observer method using inductive loops sensors.
$t_2$ and $\bar{t}$ being measured, $x_2$ is given by Equation~(\ref{eq:x_2}).
Using the same maximum likelihood estimation method, the penetration ratio $p$ is given as the
$p$ which is giving $x_2p-1+p$ as its maximum value less than $n_2$ :

\begin{equation}
\label{eq:p}
p=\frac{n_2+1}{x_2+1}
\end{equation}


It is interesting to notice that Equation~(\ref{eq:p}) is  very similar to the penetration ratio estimator derived in~\cite{9011732}.
We will show in section~\ref{simulation} that this estimator performs very well, even for low penetration ratios.
%
%
%

\section{Numerical experiments}
\label{simulation}
\begin{figure*}[htbp]
\begin{subfigure}[b]{\columnwidth}
\centering
\includegraphics[width=\columnwidth, keepaspectratio]{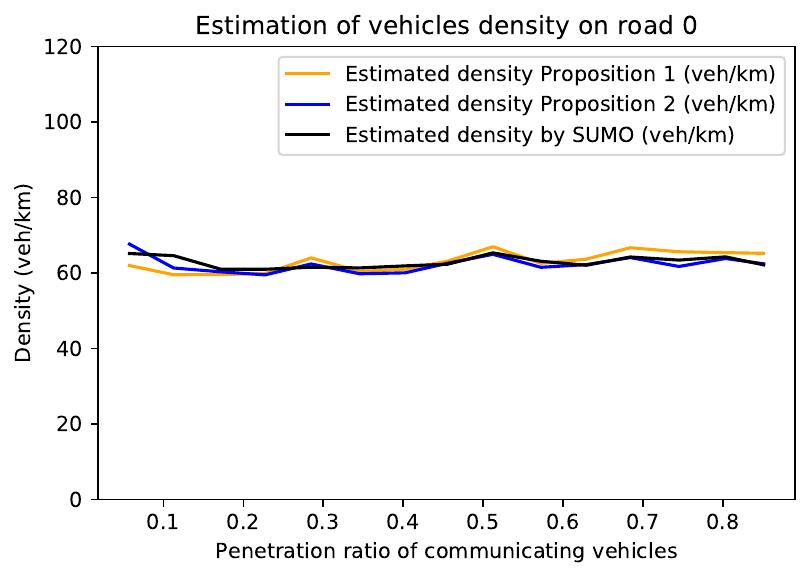}
\caption{Estimated density on road 0}
\label{fig:results0}
\end{subfigure}
\hfill
\begin{subfigure}[b]{\columnwidth}
\centering
  \includegraphics[width=\columnwidth, keepaspectratio]{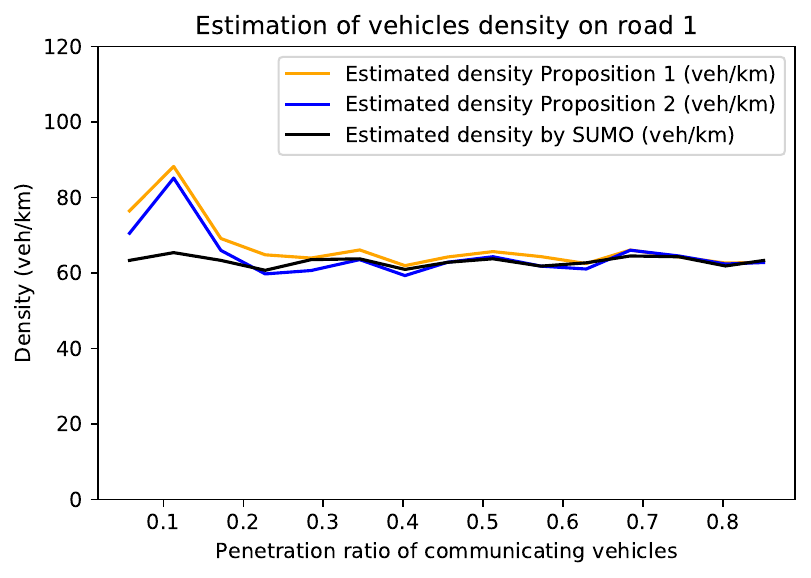}
  \caption{Estimated density on road 1}
  \label{fig:results1}
\end{subfigure}
\caption{\label{fig:results}Estimated density for roads 0 and 1.}
\end{figure*}

\begin{figure*}[htbp]
\begin{subfigure}[b]{\columnwidth}
\centering
\includegraphics[width=\columnwidth, keepaspectratio]{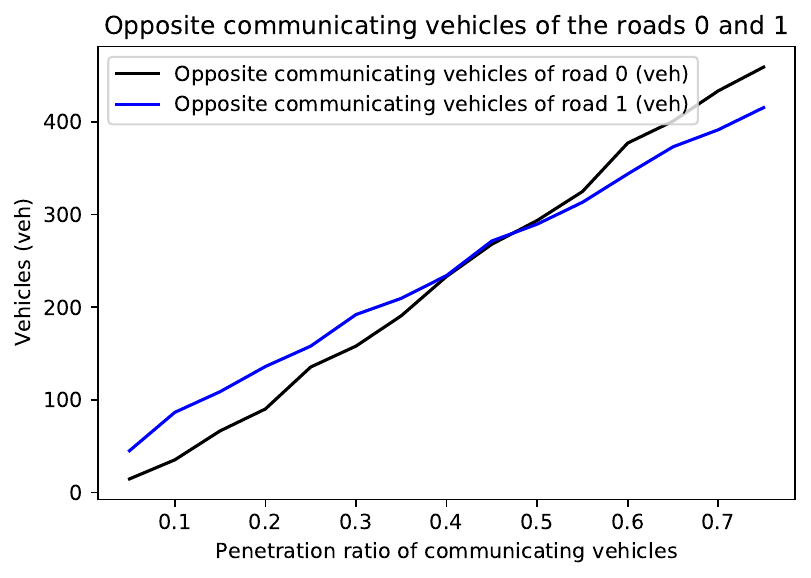}
\caption{Opposite communicating vehicles}
\label{fig:opposite}
\end{subfigure}
\hfill
\begin{subfigure}[b]{\columnwidth}
\centering
\includegraphics[width=\columnwidth, keepaspectratio]{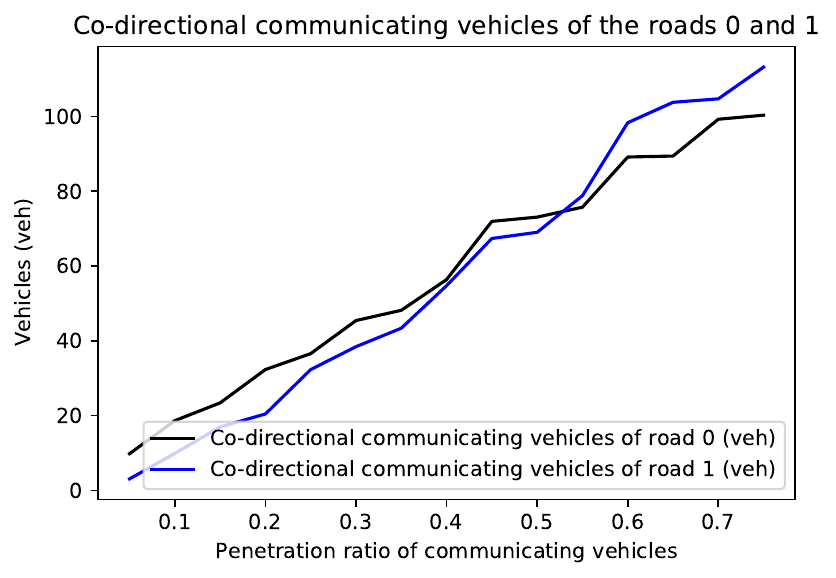}
\caption{Co-directional communicating vehicles}
\label{fig:co-directional}
\end{subfigure}
\caption{\label{fig:opp_co} Actual communicating vehicles}
\end{figure*}
\begin{figure}[htbp]
\centering
\includegraphics[width=0.9\columnwidth, keepaspectratio]{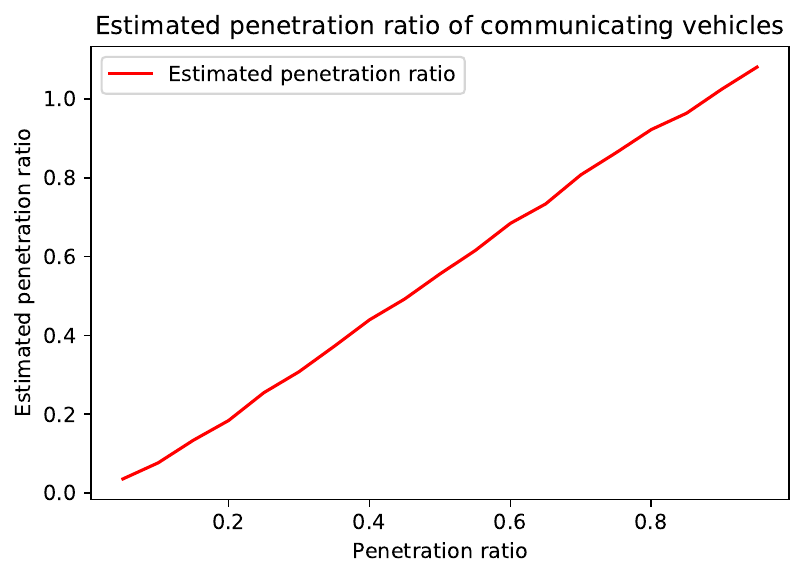}
\caption{Estimated penetration ratio of communicating vehicles}
\label{fig:results_p}
\end{figure}

\begin{figure*}[htbp]
\centering
\begin{subfigure}[b]{\columnwidth}
\begin{center}
\includegraphics[width=\columnwidth, keepaspectratio]{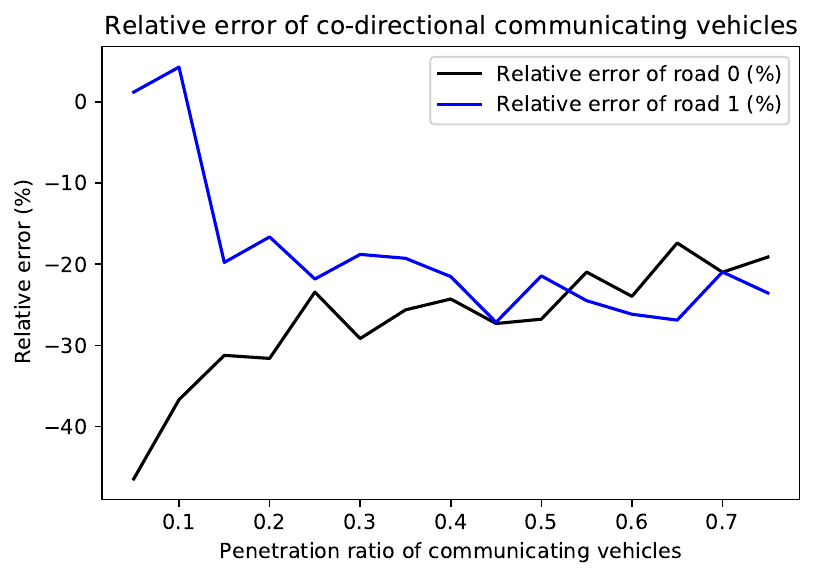}
\caption{Relative error in estimating communicating co-directional vehicles of roads 0 and 1}
\label{fig:co-directional_error}
\end{center}
\end{subfigure}
\hfill
\begin{subfigure}[b]{\columnwidth}
\begin{center}
\includegraphics[width=\columnwidth, keepaspectratio]{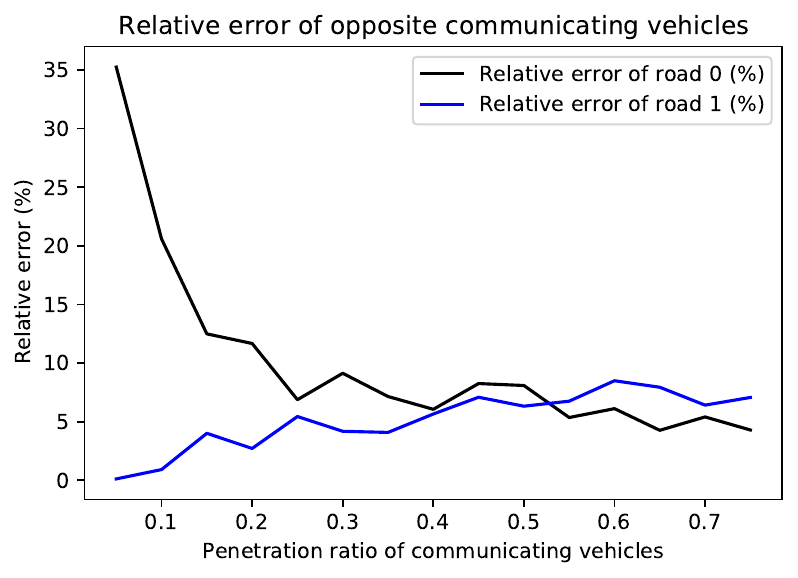}
\caption{Relative error in estimating communicating opposite vehicles of roads 0 and 1}
\label{fig:opposite_error}
\end{center}
\end{subfigure}
\hfill
\caption{Relative error in estimating communicating vehicles of roads 0 and 1}
\label{fig:error_rel}
\end{figure*}

\begin{figure*}[htbp]
\begin{subfigure}[b]{\columnwidth}
\begin{center}
\includegraphics[width=\columnwidth, keepaspectratio]{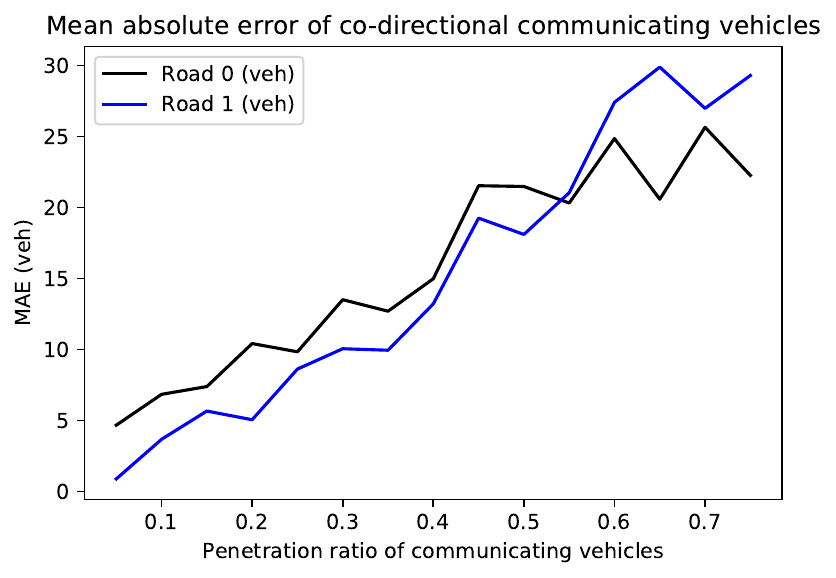}
\caption{Mean absolute error in estimating communicating co-directional vehicles of roads 0 and 1}
\label{fig:mae_co-directional_error}
\end{center}
\end{subfigure}
\hfill
\begin{subfigure}[b]{\columnwidth}
\begin{center}
\includegraphics[width=\columnwidth, keepaspectratio]{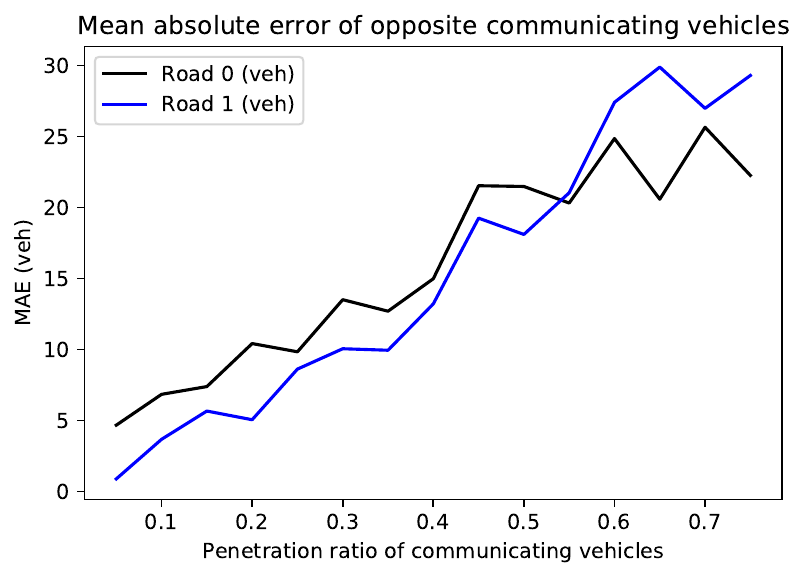}
\caption{Mean absolute error in estimating communicating opposite vehicles of roads 0 and 1}
\label{fig:mae_opposite_error}
\end{center}
\end{subfigure}
\caption{Mean absolute error in estimating communicating vehicles of roads 0 and 1}
\label{fig:error_mae}
\end{figure*}

\begin{figure*}[htbp]
\begin{subfigure}[b]{\linewidth}
\centering
\includegraphics[width=0.24\columnwidth, keepaspectratio]{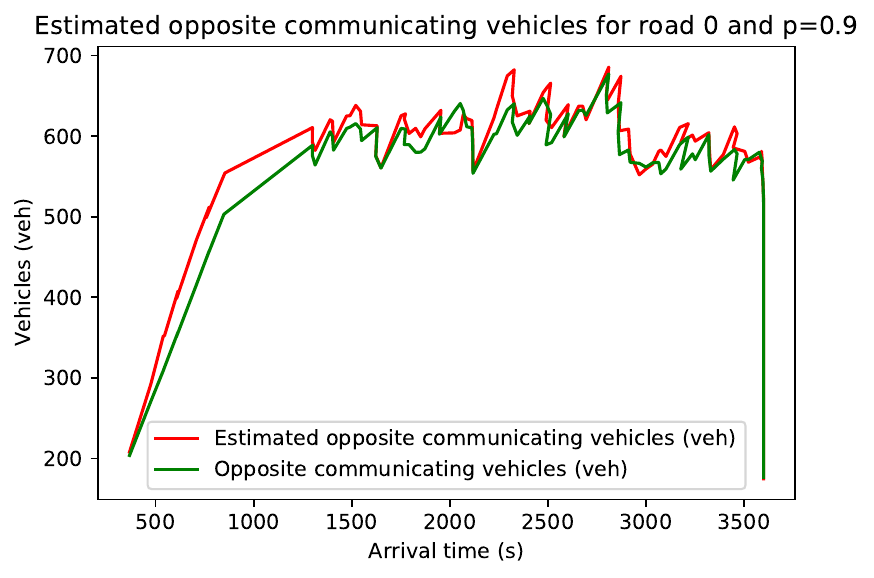} 
\includegraphics[width=0.24\columnwidth, keepaspectratio]{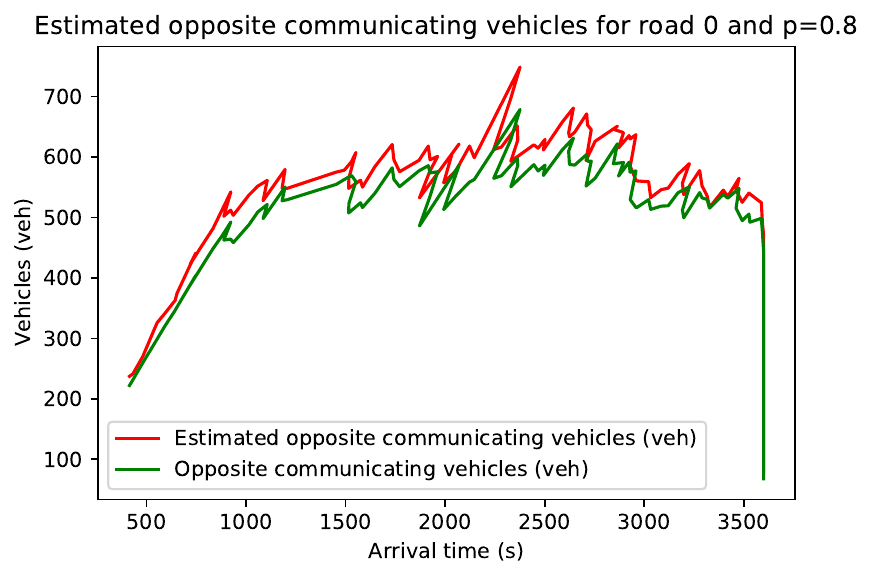} 
\includegraphics[width=0.24\columnwidth, keepaspectratio]{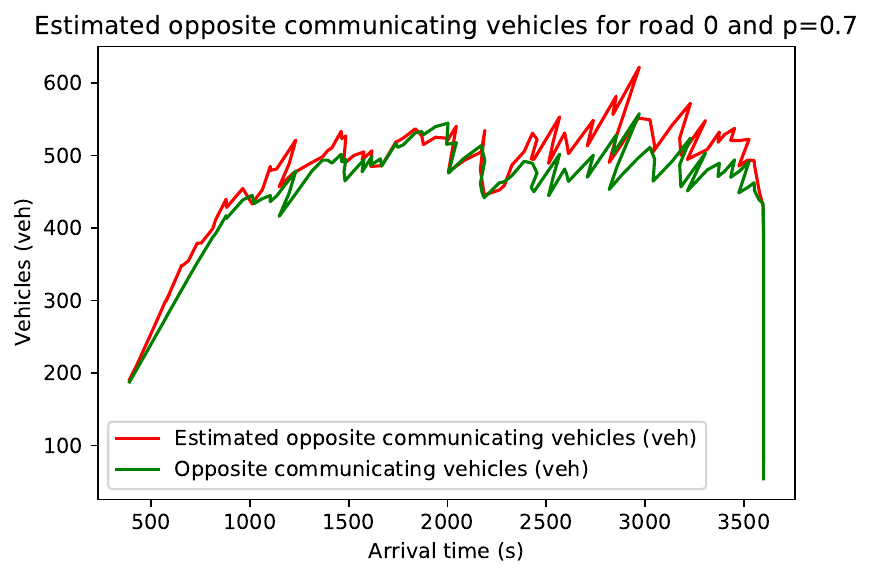} 
\includegraphics[width=0.24\columnwidth, keepaspectratio]{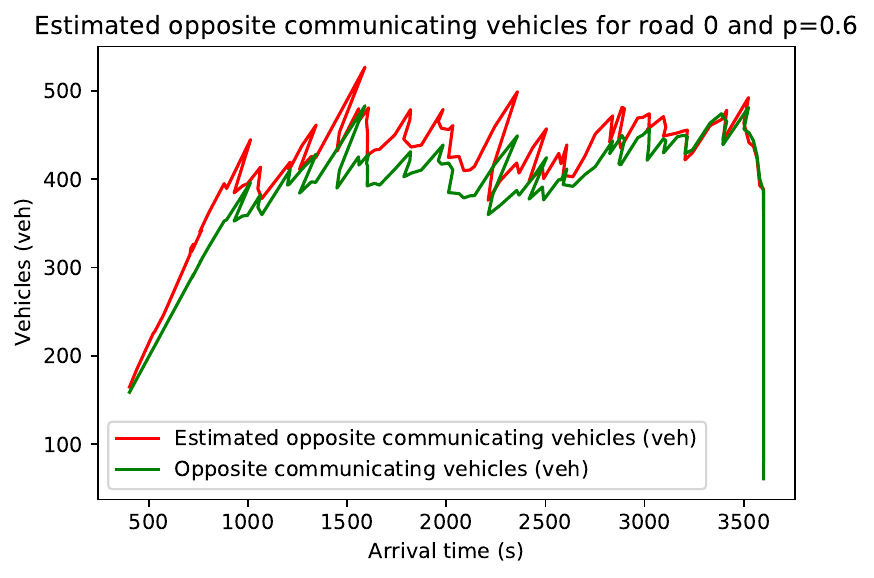} 
\end{subfigure}
\begin{subfigure}[b]{\linewidth}
\centering
\includegraphics[width=0.24\columnwidth, keepaspectratio]{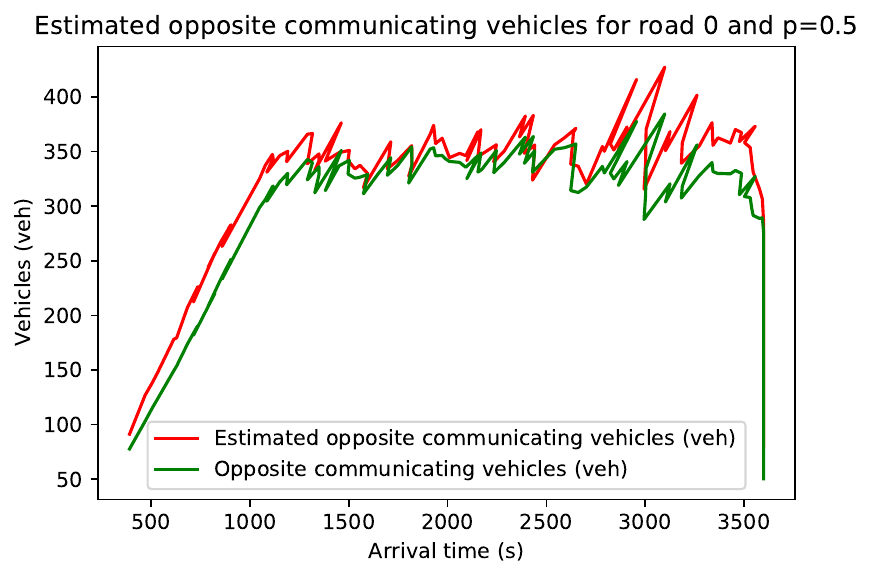} 
\includegraphics[width=0.24\columnwidth, keepaspectratio]{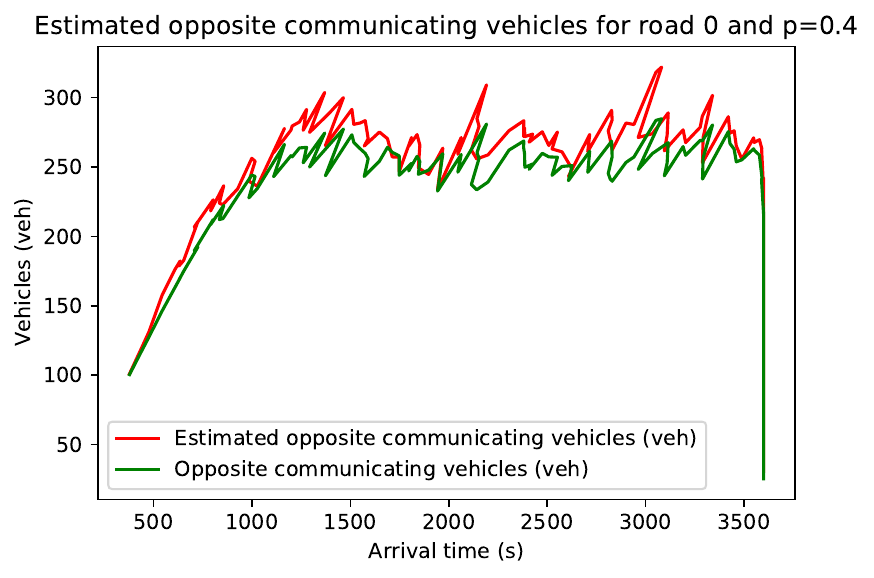} 
\includegraphics[width=0.24\columnwidth, keepaspectratio]{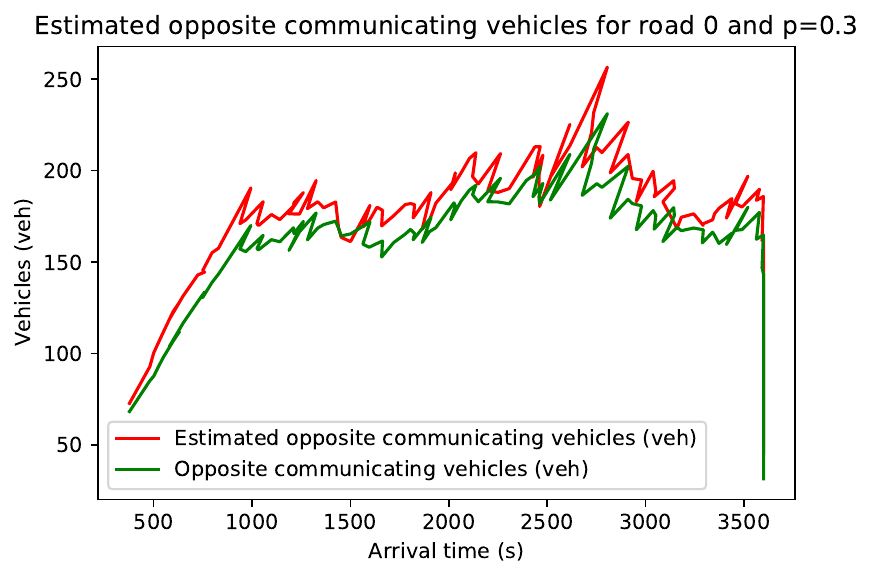} 
\includegraphics[width=0.24\columnwidth, keepaspectratio]{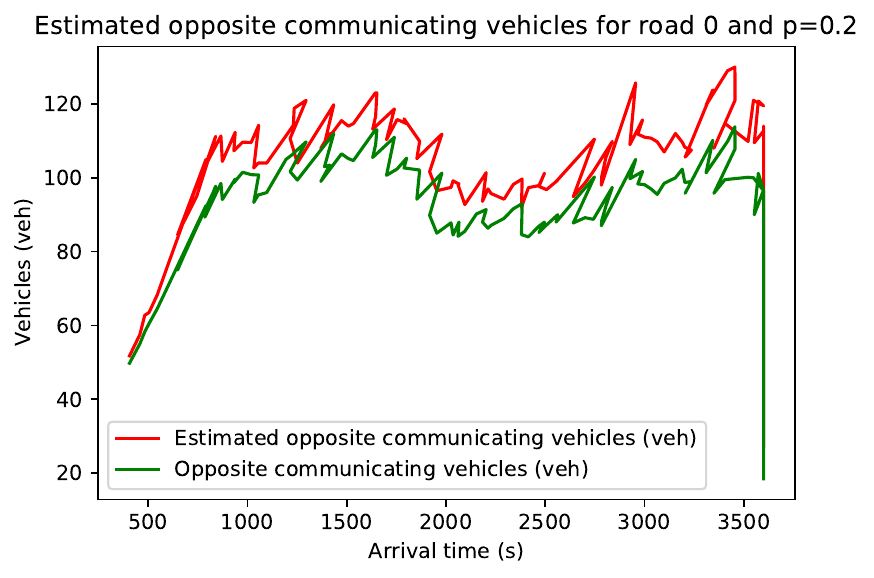} 
\end{subfigure}
\caption{Estimated communicating opposite vehicles of road 0 for various penetration ratios}
\label{fig:opposite_model}

\begin{subfigure}[b]{\linewidth}
\centering
\includegraphics[width=0.24\columnwidth, keepaspectratio]{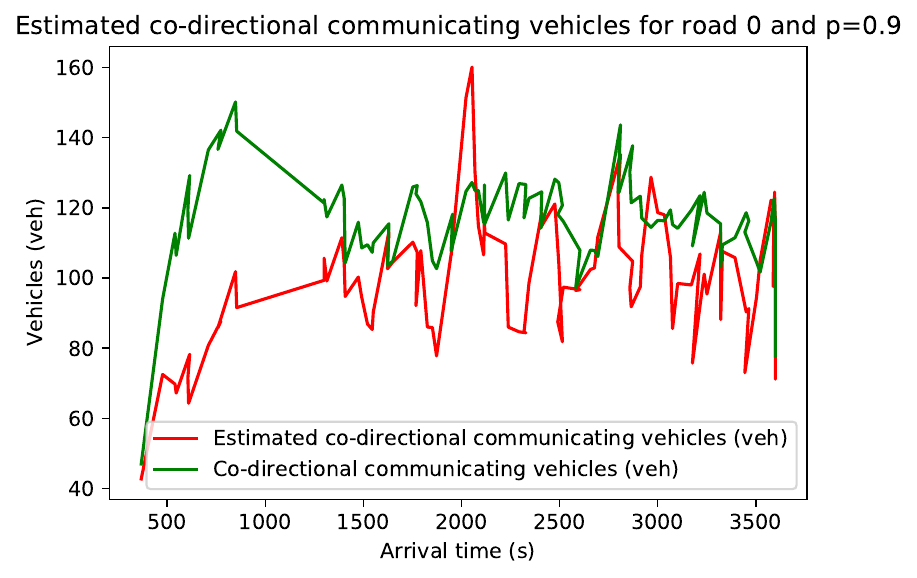} 
\includegraphics[width=0.24\columnwidth, keepaspectratio]{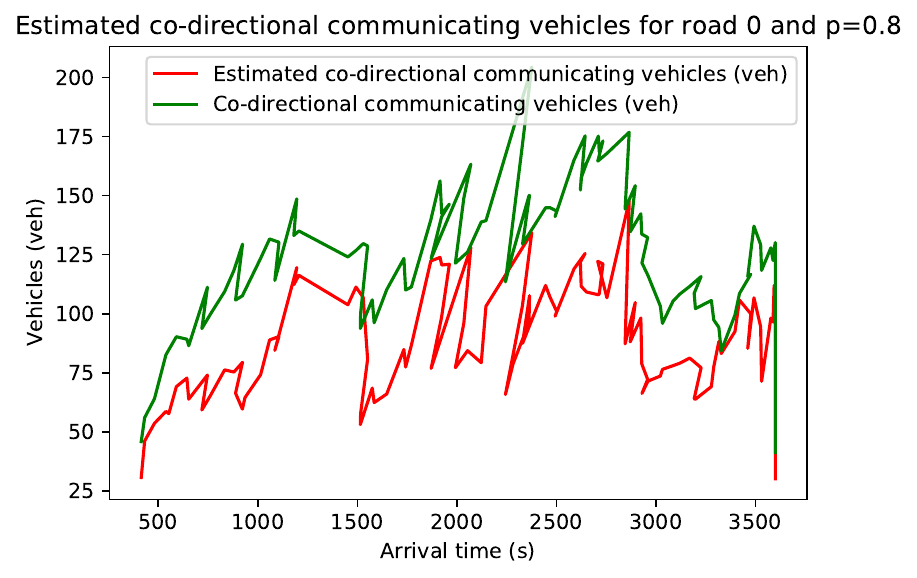} 
\includegraphics[width=0.24\columnwidth, keepaspectratio]{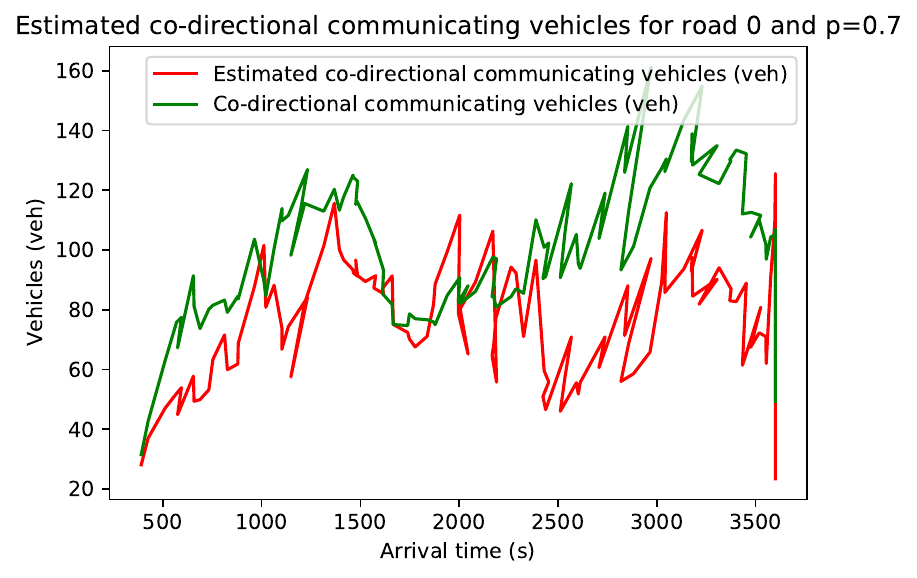} 
\includegraphics[width=0.24\columnwidth, keepaspectratio]{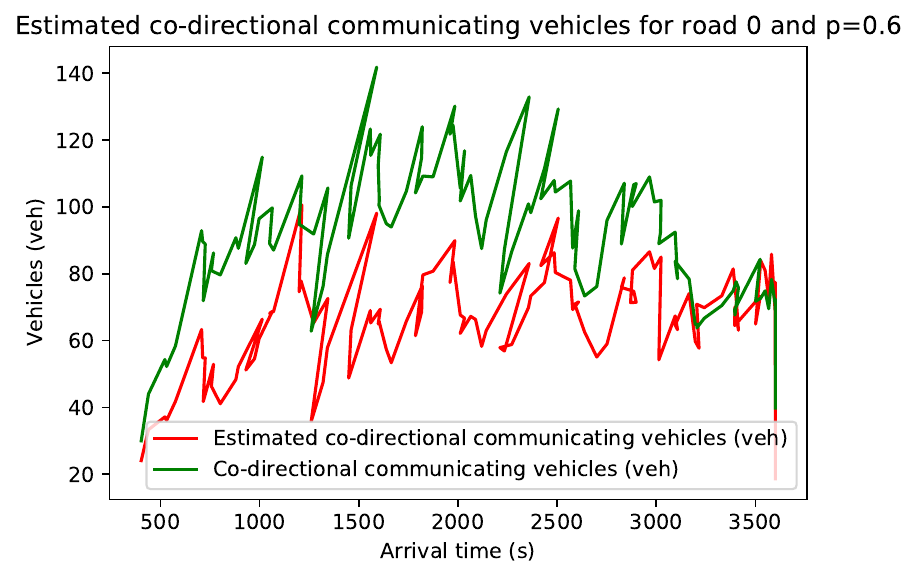} 
\end{subfigure}
\begin{subfigure}[b]{\linewidth}
\centering
\includegraphics[width=0.24\columnwidth, keepaspectratio]{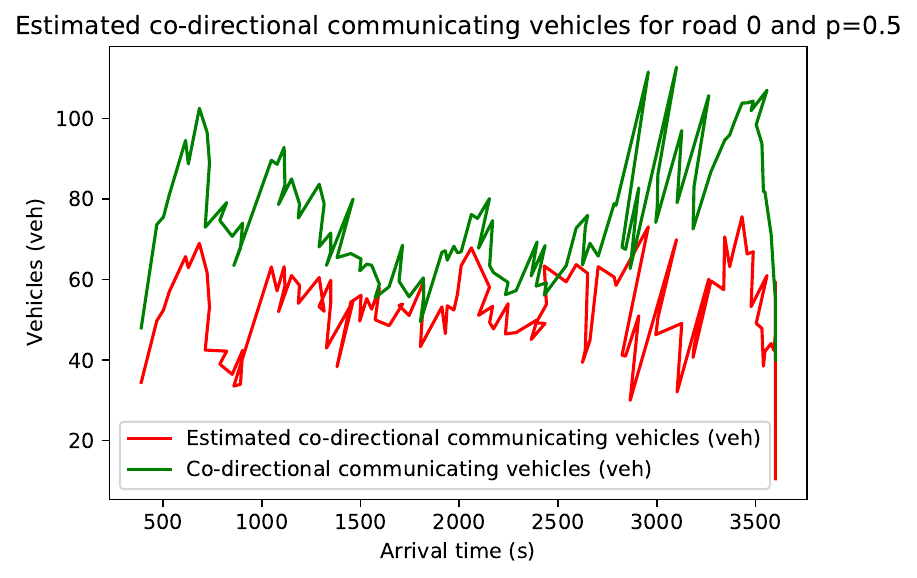} 
\includegraphics[width=0.24\columnwidth, keepaspectratio]{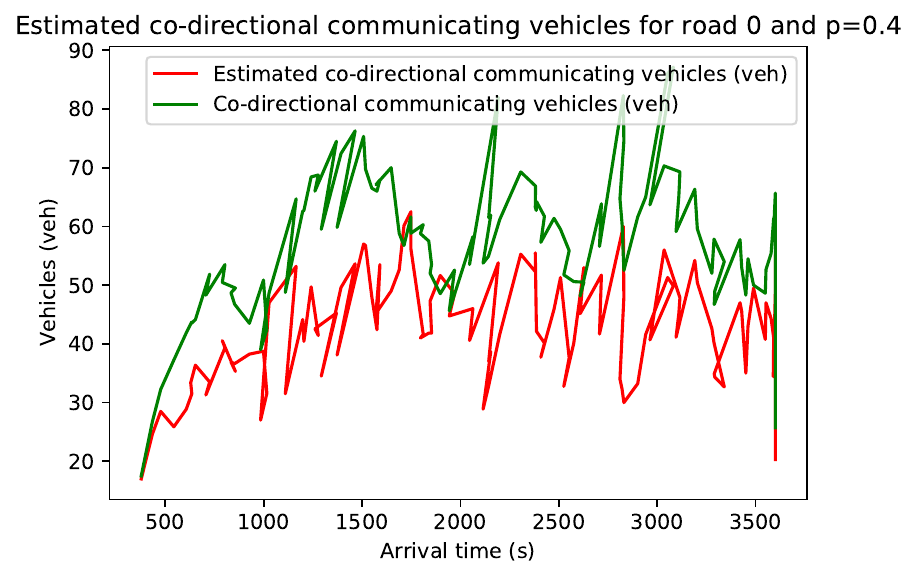} 
\includegraphics[width=0.24\columnwidth, keepaspectratio]{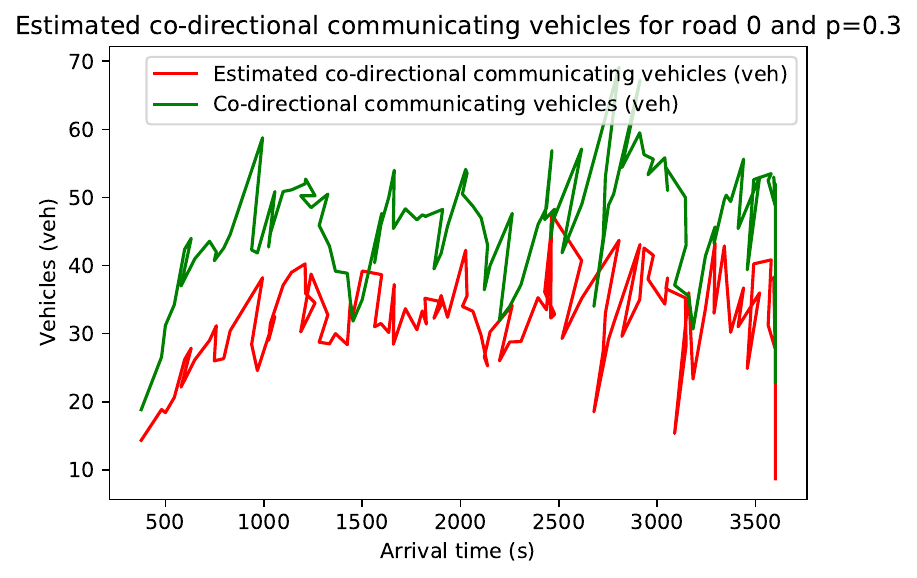} 
\includegraphics[width=0.24\columnwidth, keepaspectratio]{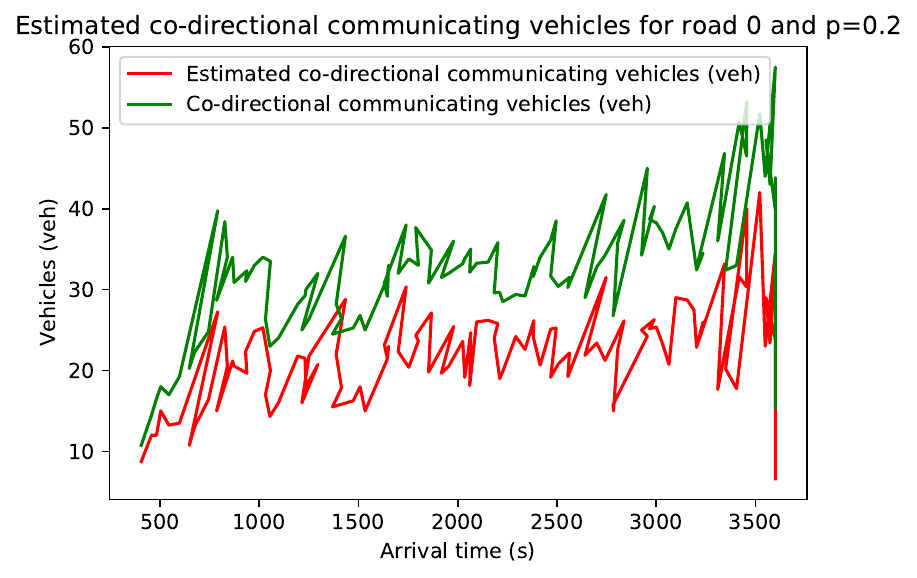} 
\end{subfigure}
\caption{Estimated communicating co-directional vehicles of road 0 for various penetration ratios}
\label{fig:codirection_model}
\end{figure*}

\subsection{Scenario and simulation tools}
We consider a two way straight road with a length of 5 km.
The two direction roads, denoted as road 0 and road 1, are composed of two lanes each.
The maximum speed of vehicles is set to $13.89$ m/s.
The road traffic demand for the simulations is given in TABLE~\ref{tab:demand}. 
The parameters for the wireless communication simulations are given in TABLE~\ref{tab:communication}.

\begin{table}[h]
\begin{center}
\caption{The road traffic demand for 3600 s simulation time.}
\begin{tabular}{|c|c|c|}
\hline
Direction & Amount of vehicles\\
\hline
West to East (Road 0) &  2400 {\it{veh/h}}\\
\hline
East to West (Road 1) &  2400 {\it{veh/h}}\\
\hline
\end{tabular}
\label{tab:demand}
\end{center}
\end{table}

\begin{table}[h]
\begin{center}
\caption{The parameters used for the IEEE 1609 wireless communication protocol simulations.}
\begin{tabular}{|c|c|c|}
\hline
Name & Value\\
\hline
Mac 1609 transmit power &  20 mW\\
\hline
Mac 1609 bitrate & 6 Mbits/s  \\
\hline
Phy 80211p min power level & -110 dBm\\
\hline
Beacon interval & 0.5 s\\
\hline
\end{tabular}
\label{tab:communication}
\end{center}
\end{table}

The simulations are performed with VEINS Framework \cite{sommer2011bidirectionally} which includes SUMO \cite{SUMO2012} as microscopic road traffic simulator and OMNET++ \cite{Varga01theomnet++} as communication simulator.
The penetration ratio of communicating vehicles is variable.
Each communicating vehicle sends periodically a beacon safety message (BSM) according to the WAVE communication protocol~\cite{6755433}.
The speed of the vehicle sending the message is included in the BSM.
The method relies on the number of BSM messages received by each communicating vehicle from its environment.
Each communicating vehicle keeps a tally of the number of BSM messages received from any other communicating vehicle in its radio range, along with the identifier of the sending vehicle, resulting in the $n_1$ and $n_2$ variables. 

The simulation results are SQL databases, processed with a Python program which applies the model to the simulation data, according to
section~\ref{model}. The main Python libraries used for post processing the simulation data results are : numpy, pandas, sqlite, and matplotlib.
The result is an estimation of the road traffic densities for each direction, aggregated over periods of time of 1 minute, which can be compared to the actual density as given by the road traffic simulator.
\subsection{Estimation of road traffic}

Fig.~\ref{fig:results0}~and~Fig.~\ref{fig:results1} display the results for the density estimations with Proposition~\ref{prop:density1}~\ref{prop:density2}.
We notice on Fig.~\ref{fig:results0}~and~Fig.~\ref{fig:results1} that the two estimations methods are close of each other, which is a sign of trust in the methods.
The asymmetry of the two roads can be explained with Fig.~\ref{fig:opposite} and Fig.~\ref{fig:co-directional}.
Indeed, we notice that the number of communicating vehicles on the road 0 is higher than the number of communicating vehicles on the road 1 for low penetration ratios.
This phenomena occurs because the algorithm which produces communicating vehicles among the total number of simulated vehicles is deterministic~\cite{sommer2011bidirectionally} and emphasises on providing an accurate penetration ratio. Because the SUMO mobility simulator demand is also deterministic, this explains the bias we can notice in these figures.

Assuming that the flow $q$ is known by another method, for example by using an inductive loop sensor, the penetration ratio of communicating vehicles is estimated.
$x_2$ is computed with Equation~(\ref{eq:x_2}) and $p$ is estimated with Equation~(\ref{eq:p}).
The results are presented in Fig.\ref{fig:results_p}.
It is interesting to notice the very good results of this estimator, even for low penetration ratios.

\subsection{Vehicles clustering}
The model is partitioning the communicating vehicles in two groups : the opposite communicating vehicles group, and the co-directional communicating vehicles group, both relative to one mobile observer. 
For each mobile observer, the number of vehicles in each group is denoted as opposite communicating vehicles, and co-directional communicating vehicles.
As the true value for the number of elements constituting these two groups is known, it is possible to evaluate the relative errors (in percentage) in estimating the sizes of the two groups.
The results are averaged on the total number of mobile observers.
In Fig.~\ref{fig:co-directional_error} and Fig.~\ref{fig:opposite_error}, we notice that the relative error for co-directional communicating vehicles group is around 25\%, which is higher than for the opposite communicating vehicles group error, which is around 5\%. These results also depend on the transmitted radio power.
This is because the total number of co-directional vehicles in the radio range is lower than the total number of opposite communicating vehicles encountered.
This number of vehicles for each group is seen on Fig.~\ref{fig:opposite} and Fig.~\ref{fig:co-directional} : the maximum number of vehicles are respectively 500 for opposite vehicles and 120 for co-directional vehicles.

The mean absolute errors for the estimation of communicating vehicles are reported in Fig.~\ref{fig:mae_co-directional_error} and Fig.~\ref{fig:mae_opposite_error}. 
These figures show that the mean absolute errors are varying between 0 and 30 vehicles, depending on the penetration ratio of communicating vehicles. The moderate error done while clustering the vehicles in two groups has little consequences on the estimation of the density of the vehicles on the roads.

In Fig.~\ref{fig:opposite_model} and Fig.~\ref{fig:codirection_model}, the number of estimated communicating vehicles for each direction, depending
on arrival times are reported. We notice in these figures that the estimation of the size of the group of opposite communicating vehicles is more accurate
than for the group of co-directional communicating vehicles. The method which counts the co-directional vehicles by using the beacon messages received
is subject to errors while communicating vehicles enter or leave the simulation. Because of that, the number of beacons may not be enough large to recognize them as
co-directional vehicles.

\section{Conclusion and perspectives}
\label{conclusion}
In this paper, the estimation of road traffic with the mobile observer method has been investigated.
It is shown that the mobile observer method can be used with only communication messages as input data.
The model has been tested on a scenario with variable penetration ratio of communicating vehicles.
The results demonstrate the potential of this approach.
As a future work, it could be interesting to test the method in the case of a congestion scenario, where the speeds of vehicles are low, and to
analyze in depth the clustering methods for differentiating the vehicles of the two roads by using messages, and especially to analyze the dependence of the
clustering of messages on the transmission power parameter.

\balance
\bibliographystyle{IEEEtran}
\bibliography{./bibliography/references.bib}

\begin{thebibliography}{10}
\providecommand{\url}[1]{#1}
\csname url@samestyle\endcsname
\providecommand{\newblock}{\relax}
\providecommand{\bibinfo}[2]{#2}
\providecommand{\BIBentrySTDinterwordspacing}{\spaceskip=0pt\relax}
\providecommand{\BIBentryALTinterwordstretchfactor}{4}
\providecommand{\BIBentryALTinterwordspacing}{\spaceskip=\fontdimen2\font plus
\BIBentryALTinterwordstretchfactor\fontdimen3\font minus
  \fontdimen4\font\relax}
\providecommand{\BIBforeignlanguage}[2]{{%
\expandafter\ifx\csname l@#1\endcsname\relax
\typeout{** WARNING: IEEEtran.bst: No hyphenation pattern has been}%
\typeout{** loaded for the language `#1'. Using the pattern for}%
\typeout{** the default language instead.}%
\else
\language=\csname l@#1\endcsname
\fi
#2}}
\providecommand{\BIBdecl}{\relax}
\BIBdecl

\bibitem{simeonova2021congestion}
E.~Simeonova, J.~Currie, P.~Nilsson, and R.~Walker, ``Congestion pricing, air
  pollution, and children’s health,'' \emph{Journal of Human Resources},
  vol.~56, no.~4, pp. 971--996, 2021.

\bibitem{GUO2019313}
\BIBentryALTinterwordspacing
Q.~Guo, L.~Li, and X.~J. Ban, ``Urban traffic signal control with connected and
  automated vehicles: A survey,'' \emph{Transportation Research Part C:
  Emerging Technologies}, vol. 101, pp. 313 -- 334, 2019. [Online]. Available:
  \url{http://www.sciencedirect.com/science/article/pii/S0968090X18311641}
\BIBentrySTDinterwordspacing

\bibitem{doi:10.1680/ipeds.1954.11628}
\BIBentryALTinterwordspacing
J.~G. WARDROP and G.~CHARLESWORTH, ``A method of estimating speed and flow of
  traffic from a moving vehicle.'' \emph{Proceedings of the Institution of
  Civil Engineers}, vol.~3, no.~1, pp. 158--171, 1954. [Online]. Available:
  \url{https://doi.org/10.1680/ipeds.1954.11628}
\BIBentrySTDinterwordspacing

\bibitem{7534864}
R.~Florin and S.~Olariu, ``On a variant of the mobile observer method,''
  \emph{IEEE Transactions on Intelligent Transportation Systems}, vol.~18,
  no.~2, pp. 441--449, 2017.

\bibitem{LEE2021100310}
\BIBentryALTinterwordspacing
M.~Lee and T.~Atkison, ``Vanet applications: Past, present, and future,''
  \emph{Vehicular Communications}, vol.~28, p. 100310, 2021. [Online].
  Available:
  \url{https://www.sciencedirect.com/science/article/pii/S2214209620300814}
\BIBentrySTDinterwordspacing

\bibitem{anwer2014survey}
M.~S. Anwer and C.~Guy, ``A survey of vanet technologies,'' \emph{Journal of
  Emerging Trends in Computing and Information Sciences}, vol.~5, no.~9, pp.
  661--671, 2014.

\bibitem{darwish_traffic_2015}
\BIBentryALTinterwordspacing
T.~Darwish and K.~Abu~Bakar, ``Traffic density estimation in vehicular ad hoc
  networks: {A} review,'' \emph{Ad Hoc Networks}, vol.~24, pp. 337--351, Jan.
  2015. [Online]. Available:
  \url{https://www.sciencedirect.com/science/article/pii/S1570870514001966}
\BIBentrySTDinterwordspacing

\bibitem{6555328}
R.~Mao and G.~Mao, ``Road traffic density estimation in vehicular networks,''
  in \emph{2013 IEEE Wireless Communications and Networking Conference (WCNC)},
  2013, pp. 4653--4658.

\bibitem{8317718}
T.~Derrmann, R.~Frank, F.~Viti, and T.~Engel, ``Estimating urban road traffic
  states using mobile network signaling data,'' in \emph{2017 IEEE 20th
  International Conference on Intelligent Transportation Systems (ITSC)}, 2017,
  pp. 1--7.

\bibitem{5073548}
D.~Valerio, A.~D'Alconzo, F.~Ricciato, and W.~Wiedermann, ``Exploiting cellular
  networks for road traffic estimation: A survey and a research roadmap,'' in
  \emph{VTC Spring 2009 - IEEE 69th Vehicular Technology Conference}, 2009, pp.
  1--5.

\bibitem{artimy_local_2007}
\BIBentryALTinterwordspacing
M.~Artimy, ``Local {Density} {Estimation} and {Dynamic} {Transmission}-{Range}
  {Assignment} in {Vehicular} {Ad} {Hoc} {Networks},'' \emph{IEEE Transactions
  on Intelligent Transportation Systems}, vol.~8, no.~3, pp. 400--412, Sep.
  2007, conference Name: IEEE Transactions on Intelligent Transportation
  Systems. [Online]. Available:
  \url{https://ieeexplore.ieee.org/document/4298893}
\BIBentrySTDinterwordspacing

\bibitem{yu_vanet_2013}
\BIBentryALTinterwordspacing
H.~Yu, J.~Yoo, and S.~Ahn, ``A {VANET} routing based on the real-time road
  vehicle density in the city environment,'' in \emph{2013 {Fifth}
  {International} {Conference} on {Ubiquitous} and {Future} {Networks}
  ({ICUFN})}, Jul. 2013, pp. 333--337, iSSN: 2165-8536. [Online]. Available:
  \url{https://ieeexplore.ieee.org/document/6614836}
\BIBentrySTDinterwordspacing

\bibitem{seo_traffic_2017}
\BIBentryALTinterwordspacing
T.~Seo, A.~M. Bayen, T.~Kusakabe, and Y.~Asakura,
  ``\BIBforeignlanguage{en}{Traffic state estimation on highway: {A}
  comprehensive survey},'' \emph{\BIBforeignlanguage{en}{Annual Reviews in
  Control}}, vol.~43, pp. 128--151, Jan. 2017. [Online]. Available:
  \url{https://www.sciencedirect.com/science/article/pii/S1367578817300226}
\BIBentrySTDinterwordspacing

\bibitem{bauza_traffic_2013}
\BIBentryALTinterwordspacing
R.~Bauza and J.~Gozalvez, ``Traffic congestion detection in large-scale
  scenarios using vehicle-to-vehicle communications,'' \emph{Journal of Network
  and Computer Applications}, vol.~36, no.~5, pp. 1295--1307, Sep. 2013.
  [Online]. Available:
  \url{https://www.sciencedirect.com/science/article/pii/S1084804512000628}
\BIBentrySTDinterwordspacing

\bibitem{henderson2008network}
T.~R. Henderson, M.~Lacage, G.~F. Riley, C.~Dowell, and J.~Kopena, ``Network
  simulations with the ns-3 simulator,'' \emph{SIGCOMM demonstration}, vol.~14,
  no.~14, p. 527, 2008.

\bibitem{9807677}
R.~Florin and S.~Olariu, ``Real-time traffic density estimation: Putting
  on-coming traffic to work,'' \emph{IEEE Transactions on Intelligent
  Transportation Systems}, vol.~24, no.~1, pp. 1374--1383, 2023.

\bibitem{krajzewicz_recent_2012}
D.~Krajzewicz, J.~Erdmann, M.~Behrisch, and L.~Bieker,
  ``\BIBforeignlanguage{en}{Recent {Development} and {Applications} of {SUMO}
  – {Simulation} of {Urban} {MObility}},'' 2012.

\bibitem{florin_towards_2020}
\BIBentryALTinterwordspacing
R.~Florin and S.~Olariu, ``\BIBforeignlanguage{en}{Towards real-time density
  estimation using vehicle-to-vehicle communications},''
  \emph{\BIBforeignlanguage{en}{Transportation Research Part B:
  Methodological}}, vol. 138, pp. 435--456, Aug. 2020. [Online]. Available:
  \url{https://www.sciencedirect.com/science/article/pii/S0191261520303404}
\BIBentrySTDinterwordspacing

\bibitem{kuwahara_traffic_2021}
\BIBentryALTinterwordspacing
M.~Kuwahara, A.~Takenouchi, and K.~Kawai, ``\BIBforeignlanguage{en}{Traffic
  state estimation by backward moving observers: {An} application and
  validation under an incident},'' \emph{\BIBforeignlanguage{en}{Transportation
  Research Part C: Emerging Technologies}}, vol. 127, p. 103158, Jun. 2021.
  [Online]. Available:
  \url{https://www.sciencedirect.com/science/article/pii/S0968090X21001765}
\BIBentrySTDinterwordspacing

\bibitem{VANERP2018281}
\BIBentryALTinterwordspacing
P.~B. {van Erp}, V.~L. Knoop, and S.~P. Hoogendoorn, ``Macroscopic traffic
  state estimation using relative flows from stationary and moving observers,''
  \emph{Transportation Research Part B: Methodological}, vol. 114, pp.
  281--299, 2018. [Online]. Available:
  \url{https://www.sciencedirect.com/science/article/pii/S0191261518302285}
\BIBentrySTDinterwordspacing

\bibitem{9864074}
C.~Nguyen Van~Phu and N.~Farhi, ``Estimation of road traffic state at a
  multilanes controlled junction,'' \emph{IEEE Transactions on Intelligent
  Transportation Systems}, vol.~23, no.~12, pp. 23\,657--23\,667, 2022.

\bibitem{comert_queue_2016}
\BIBentryALTinterwordspacing
G.~Comert, ``Queue length estimation from probe vehicles at isolated
  intersections: {Estimators} for primary parameters,'' \emph{European Journal
  of Operational Research}, vol. 252, no.~2, pp. 502 -- 521, 2016. [Online].
  Available:
  \url{http://www.sciencedirect.com/science/article/pii/S0377221716000850}
\BIBentrySTDinterwordspacing

\bibitem{wardrop_road_1952}
\BIBentryALTinterwordspacing
J.~G. Wardrop, ``Road paper. some theoretical aspects of road traffic
  research.'' \emph{Proceedings of the Institution of Civil Engineers}, vol.~1,
  no.~3, pp. 325--362, May 1952, publisher: ICE Publishing. [Online].
  Available:
  \url{https://www.icevirtuallibrary.com/doi/abs/10.1680/ipeds.1952.11259}
\BIBentrySTDinterwordspacing

\bibitem{6755433}
``Ieee guide for wireless access in vehicular environments (wave) -
  architecture,'' \emph{IEEE Std 1609.0-2013}, pp. 1--78, March 2014.

\bibitem{sommer2011bidirectionally}
C.~Sommer, R.~German, and F.~Dressler, ``{Bidirectionally Coupled Network and
  Road Traffic Simulation for Improved IVC Analysis},'' \emph{IEEE Transactions
  on Mobile Computing}, vol.~10, no.~1, p. 3–15, January 2011.

\bibitem{varga_overview_2010}
\BIBentryALTinterwordspacing
A.~Varga and R.~Hornig, ``{AN} {OVERVIEW} {OF} {THE} {OMNeT}++ {SIMULATION}
  {ENVIRONMENT},'' May 2010. [Online]. Available:
  \url{https://eudl.eu/doi/10.4108/icst.simutools2008.3027}
\BIBentrySTDinterwordspacing

\bibitem{turner2017using}
S.~Turner, P.~Koeneman \emph{et~al.}, ``Using mobile device samples to estimate
  traffic volumes,'' Minnesota. Dept. of Transportation. Research Services \&
  Library, Tech. Rep., 2017.

\bibitem{LIU201683}
\BIBentryALTinterwordspacing
J.~Liu and A.~J. Khattak, ``Delivering improved alerts, warnings, and control
  assistance using basic safety messages transmitted between connected
  vehicles,'' \emph{Transportation Research Part C: Emerging Technologies},
  vol.~68, pp. 83--100, 2016. [Online]. Available:
  \url{https://www.sciencedirect.com/science/article/pii/S0968090X1630002X}
\BIBentrySTDinterwordspacing

\bibitem{myung_tutorial_2003}
\BIBentryALTinterwordspacing
I.~J. Myung, ``\BIBforeignlanguage{en}{Tutorial on maximum likelihood
  estimation},'' \emph{\BIBforeignlanguage{en}{Journal of Mathematical
  Psychology}}, vol.~47, no.~1, pp. 90--100, Feb. 2003. [Online]. Available:
  \url{https://www.sciencedirect.com/science/article/pii/S0022249602000287}
\BIBentrySTDinterwordspacing

\bibitem{binomial_rundel}
C.~Rundel, ``{Lecture 5: Binomial Distribution},''
  \url{https://www2.stat.duke.edu/courses/Spring12/sta104.1/Lectures/Lec5.pdf},
  2012.

\bibitem{9011732}
C.~Nguyen Van~Phu and N.~Farhi, ``Estimation of urban traffic state with probe
  vehicles,'' \emph{IEEE Transactions on Intelligent Transportation Systems},
  vol.~22, no.~5, pp. 2797--2808, 2021.

\bibitem{SUMO2012}
D.~Krajzewicz, J.~Erdmann, M.~Behrisch, and L.~Bieker, ``Recent development and
  applications of {SUMO - Simulation of Urban MObility},'' \emph{International
  Journal On Advances in Systems and Measurements}, vol.~5, no. 3\&4, pp.
  128--138, December 2012.

\bibitem{Varga01theomnet++}
A.~Varga, ``The omnet++ discrete event simulation system,'' in \emph{In
  ESM’01}, 2001.

\end{thebibliography}
\end{document}